\UseRawInputEncoding
\documentclass[aps,titlepage,12pt,superscriptaddress]{revtex4}
\usepackage{epsfig}
\usepackage{times,epstopdf}
\begin{document}

\title{Dynamics of the COVID-19 pandemic in India}

\author{Subhas Khajanchi}
\email{subhas.maths@presiuniv.ac.in}
\affiliation{Department of Mathematics, Presidency University, 86/1 College Street, Kolkata 700073, India}

\author{Kankan Sarkar}
\affiliation{Department of Mathematics, Malda College, Malda, West Bengal 712101, India}

\author{Jayanta Mondal}
\affiliation{Department of Mathematics, Diamond Harbour Women's University, Diamond Harbour Road, Sarisha 743368, India}

\begin{abstract}\noindent
\\ Understanding the dynamics of the COVID-19 pandemic is crucial for improved control and social distancing strategies. To that effect, we have employed the susceptible-exposed-infectious-recovered model, refined by contact tracing and hospitalization data from Indian provinces Kerala, Delhi, Maharashtra, and West Bengal, as well as from overall India. We have performed a sensitivity analysis to identify the most crucial input parameters, and we have calibrated the model to describe the data as best as possible. Short-term predictions reveal an increasing and worrying trend of COVID-19 cases for all four provinces and India as a whole, while long-term predictions also reveal the possibility of oscillatory dynamics. Our research thus leaves the option open that COVID-19 might become a seasonal occurrence. We also simulate and discuss the impact of media on the dynamics of the COVID-19 pandemic.
\end{abstract}

\maketitle

\section{Introduction}

\noindent According to the World Heath Organization (WHO), 1,878,489 confirmed cases (119,044 confirmed deaths) of novel coronavirus disease 2019 (2019-nCoV) has been reported throughout the world, as of April 15, 2020, including 11,439 confirmed cases (377 deaths) from India \cite{A-who}. The outbreak was first announced by the ``Health Commission of Hubei province'', China, a cluster of unexplained cases of pneumonia of unknown etiology (unknown cases) \cite{Zhu20}, which is lethal, was first identified in Wuhan, the capital city of Hubei and seven largest city of China, on December 31, 2019 \cite{Chan20,Cohen20,Tang20}. Later on, the novel coronavirus disease was called officially as COVID-19 by WHO \cite{A-who}. The epidemic was announced a major public concern worldwide on January 30, 2020 by the World Health Organization \cite{Zhu20}. Coronavirus, an enveloped virus described by a single-stranded, positive-sense RNA belonging to the family of Coronaviridae and the Nidovirales order and extensively disseminated among humans as well as mammals \cite{Chen20}. The ongoing coronavirus epidemic has been announced a pandemic by the World Health Organization (WHO) on January 30, 2020, and the Govt. of India has announced 21 days nationwide lockdown from March 25, 2020  to April 14, 2020, to prevent stage-III spreading of the virus or human-to-human transmission.

Coronavirus causes typically mild infections but sometimes lethal communicable disorders leading to Severe Acute Respiratory Syndrome (SARS) and Middle East Respiratory Syndrome (MERS) and in present COVID-19 pandemic, and in general it can be thought as SARSr-CoV \cite{Gumel04}, which is mostly observed in bats-could create a future disease outbreak \cite{Drosten03,Zhou20}. The most important modes of transmission of coronavirus are respiratory droplets and contact transmission (contaminated hands) and it has incubation period 2-14 days \cite{AA-who,Chan20}. From the confirmed cases of coronavirus, the symptoms range from fever, sneeze, or even a runny nose, dry cough fatigue breathing problem and lung infiltration to severely ill and dying \cite{AA-who}. Shared surfaces are also an important risky since the virus clings on to surfaces and propagates as soon as another person touches the shared surface. The virus also can remain active for about 3 hours after aerosolization (droplets carrying virus mixed with air), on plastic for fifteen hours, and on metallic surfaces for five-six hours.

The continuing coronavirus outbreak has been announced a widespread by the World Health Organization, as of March 16, 2020, the United Nations' Public Health Agency has authenticated 1,67,511 confirmed cases, including 6,606 deaths across 151 countries throughout the world \cite{A-who,Chan20}. As countries start shutting down borders, cutting off international marketing and quarantining their individuals, the key to safeguard our families, near ones and communities lies in understanding the nature and behavior and why and how this particular virus is developing. Some other factors like population mobility by air travel, the documented human-to-human transmission, low air temperature and low humidity highly affect the transmission of coronavirus  \cite{A-who,Jingyuan20}.

Throughout the world, the air travel is one of the most important dynamic network, and India is in 17th position worldwide the countries with highest chance of importation of coronavirus through air traveler \cite{AA-trav,Mandal20}. The probability of an infected air traveler to reach to India at final destination is 0.383\% (as on April 07, 2020), with maximum relative import risk in Delhi (0.124\%, as on April 07, 2020), Mumbai (0.064\%, as on April 07, 2020), Kolkata (0.021\%, as on April 07, 2020), Madras, Kochi, Bangalore, Hyderabad and so on \cite{AA-trav}.

At the beginning, the Ministry of Health and Family Welfare (MoHFW), Govt. of India suggested to discontinue to travel from China to India and send to quarantine who are coming from China \cite{AA-CH-tra}. Those who return from Wuhan, the sprawling capital city of Hubei province in China, later on January 15, 2020 were to be examined for coronavirus. Also, those who are feeling sick within one month of returning from China were suggested to admit the nearby health center in addition to maintain self-isolation at home \cite{AA-CH-tra1}. At the very beginning, thermal entry screening of travelers from Wuhan (China) was set up around twenty one airports throughout the country with international screening for flights from Tiwan, China, Singapore, South Korea, Italy, Hong-Kong, Japan and so on. Symptomatic ill travelers are suggested to volunteer for screening test. Similar screening test has also be taken into account at the international harbor \cite{AA-CovIn}.

Governments will not be competent to diminish both fatalities from coronavirus epidemic and the economic impact of viral outbreak. Maintaining the fatality rate as minimum as possible will be the utmost importance for the populations; therefore the Governments must put in place measures to mitigate the unavoidable economic downturn. In our viewpoint, coronavirus has turned into a pandemic, with small chains of transmission in several countries and big chains of transmission resulting in large-scale outbreak in most of the countries, namely United States, Spain, Italy, South Korea, Germany, France and so on \cite{A-who}.

Till date, there is no specific licensed vaccine, antivirals or effective therapeutics to treat coronavirus infections. Due to absence of coronavirus therapeutics, we utilized the Non-Pharmacological interventions (NPIs) - focused at minimizing transmission by reducing contact rates among individuals. As for examples the measures adopted in this time incorporated social distancing, closing schools, universities, offices, churches, bars, avoid mass gatherings,  other social places as well as contacts of cases (quarantine, surveillance, contact tracing) \cite{Ferguson20}.

The route of an outbreak can be described by a series of important factors, but some of which are extremely difficult to understand at present coronavirus disease. The basic reproduction number $R_0$ is one of the most crucial quantities in infectious diseases, as  $R_0$ measures how contagious a disease is. $R_0$ designates the number of secondary infections infected by one infected individual in a whole susceptible class, or more specifically the area under epidemic curve. For $R_0 < 1$, the disease is expected to stop spreading, but for  $R_0 = 1$ an infected individual can infect on an average 1 person, that is, the spread of the disease is stable.  The disease can spread and become epidemic if $R_0$ must be greater than 1. $R_0$ can help us to understand the effectiveness of the disease, that is, under what condition the disease  can stop or spread ? The values for $R_0$ in China is around 2.5 in the primary stage of the outbreak of coronavirus \cite{Anderson20}.

The spread of coronavirus outbreak depends on the infectivity of the virus  and the availability of susceptible individuals. Mathematical modeling play a vital role to better understand the disease dynamics and designing policies to manage quickly spreading infectious diseases in lack of effective vaccine or specific antivirals \cite{Egger17}. Recently, many mathematical model have already been studied to understand the complicated dynamics of novel coronavirus, and some of these are cited in our references \cite{Chan20,Chen20a,Imai20,Liu20,Mandal20,Nadim20,Tang20,Tang20a,Wu20}. A novel coronavirus model has been established by Chen et al. \cite{Chen20a} and find the basic reproduction number. Imai et al. \cite{Imai20} studied a computational model for novel coronavirus disease 2019 in Wuhan, a city of China and their model mainly focused on human-to-human transmission. Tang et al. \cite{Tang20} established a compartmental model for coronavirus by incorporating symptomatic ill class to obtain the patients' epidemiological status. They obtained the basic reproduction number 6.47, which is very high for the infectious diseases. Nadim et al. \cite{Nadim20} studied a mathematical to investigate the coronavirus disease, where they performed stability analysis and they validated their model with the data from Hubei, the city of China. Their model is originally established by Gumel et al. \cite{Gumel04} for Severe Acute Respiratory Syndrome (SARS) outbreak. Wu et al. \cite{Wu20} established a coronavirus model by considering four classes, namely susceptible, exposed, infected and recovered to study the human-human transmission dynamics based on the data from December 31, 2019 to January 28, 2020. They computed the reproduction number approximately 2.68 for coronavirus disease.

The conventional models governing the outbreak of infectious diseases mainly depend on the interplays among susceptible, infected and exposed classes. Moreover, the other aspects, such as vaccination, huge news coverage, educational campaigns and rapid information flow can create great psychological effects on the population, and thus remarkably change the publics' behavior and affect the implementation of individuals' intervention and control strategies \cite{Funk09}. It is mainly the awareness program due to media which make the individuals enlighten regarding the disease to take safeguards such as vaccination, wearing protective masks, social distancing etc., to minimize their risks of being infected \cite{d'Onofrio07, d'Onofrio09}. How long and how effectual media related impact persists is thus an important issue for future outbreak, and quantifying this impact through a  modeling process falls within the scope of our investigation. Several mathematical models have been incorporated with the assumption that the media related awareness has an impact on reducing the contact rate of susceptible and infected population \cite{Buonomo08,Funk09,Kumar17,Sun11,Das20,Xiao15}. Influenced by those literatures we develop and analyze a mathematical model for coronavirus to study the impact of media related awareness program.

To obtain a better insight into the important factors related with the control of novel coronavirus in a community and throughout the world, we investigate a dynamic model to study  the epidemic of coronavirus and its control in four mostly affected states, namely Maharashtra, Delhi, Kerala, West Bengal and the overall India.
We estimated most important parameters, namely $\beta_s$, $\xi_a$, $q_a$, $\alpha_h$ $\xi_h$, $\xi_i$ and $\gamma_i$ by using least square method. We calculate the basic reproduction for our model for four different states as well as the India.

\section{Dynamic model without effective control measures}

\noindent Here, we extend the classical deterministic susceptible-exposed-infectious-removed (SEIR) compartmental model refined by introducing contact tracing-hospitalization strategies to study the epidemiological properties of coronavirus (COVID-19). We calibrate our mathematical model using data gained form confirmed cases of coronavirus in India and estimated the basic reproduction number for the disease transmission. In order to make our mathematical model more realistic, we consider the following assumptions:
\begin{itemize}
  \item the model involves a net inflow rate of susceptible individuals $\Lambda_{s}$ per unit time,
  \item the model has no zoonotic infections of coronavirus, considering only the epidemic spread through human beings,
  \item there is no effective control measures before April 10, 2020,
  \item the model involves some demographic effects by accepting a proportional natural mortality in each of the sub-populations.
\end{itemize}

In the underlying dynamic model, the total population $N(t)$ is classified into six sub-populations (classes), namely susceptible $S(t)$, exposed $E(t)$, asymptomatic $A(t)$, clinically ill or symptomatic $I(t)$, hospitalized $H(t)$ and recovered $R(t)$, respectively.

\begin{figure}
  \centering
  \includegraphics[width=1.0\textwidth]{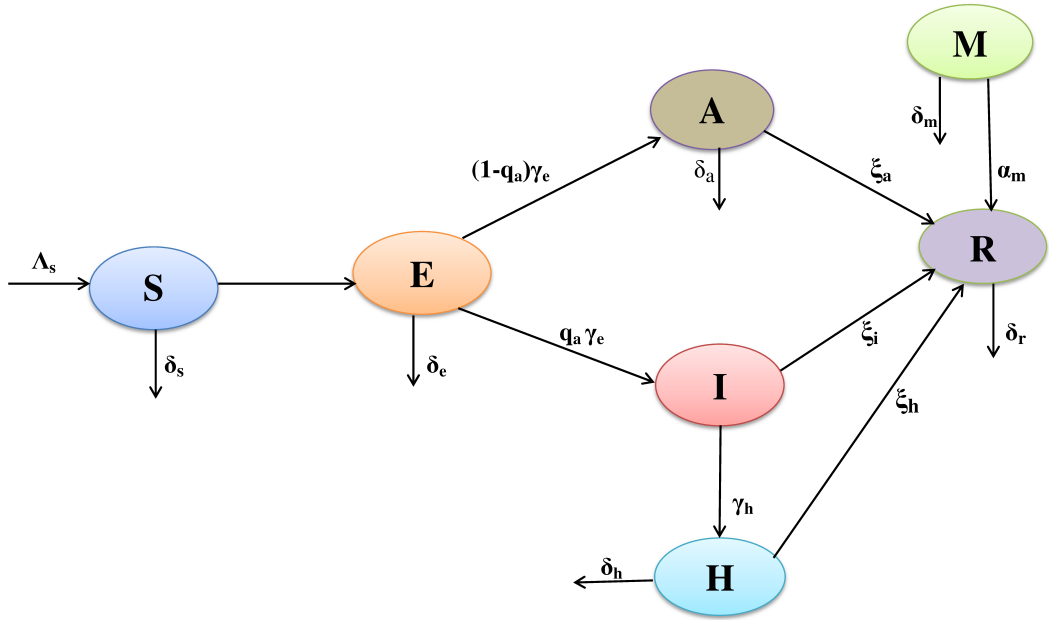} \\
  \caption{ The schematic flow diagram represents the biological mechanism of novel coronavirus (COVID-19) infection in India, which influences the formulation of the mathematical model (\ref{stateeq}). The mathematical model consists of seven sub-populations including media related information index: susceptible $S(t)$, exposed $E(t)$, asymptomatic $A(t)$, symptomatic or clinically ill $I(t)$, hospitalized or isolated $H(t)$, recovered $R(t)$ and media awareness $M(t)$ individuals in a total population of $N(t)$ = $S(t)$ + $E(t)$ + $A(t)$ + $I(t)$ + $H(t)$ + $R(t)$ individuals. }
\label{Schema}
\end{figure}

The coronavirus outbreaks in the population throughout the world and so the news concerning coronavirus epidemics through different media including television, social networking sites and newspaper as well as educational programs from government, non-government organization and local bodies. The source of information density of media is assumed to be proportional to the number of symptomatic/infective individuals and will alter as the symptomatic population alters. Herein, $M(t)$ denotes the cumulative density of media in individual such that $M(t)$ = 0, for $I(t)$ = 0. This is mimic to the mathematical model developed by d'Onofrio et al. \cite{d'Onofrio07}, in the case of non-delayed model and mimics to the media variable as incorporated there. This media creates the behavioral alteration in the susceptible populations to save themselves from contracting infection. As for example, Government of Senegal has taken an initiative through SMS campaign to spread the information regarding Ebola awareness \cite{Govt}. Albeit individuals are informed through media, not everyone responses to it due to insufficiency of resources, irresponsive nature and economic condition etc. Therefore a part of susceptible individuals with news regarding coronavirus, responding to the news for corona symptoms and altering their behavior to avoid infection, and will move to the recovered individuals. The rate of behavioral response through media has been considered to be a function of both the densities for susceptible populations and media, that is, $f(S(t), M(t))$. Moreover, we assume that the source of media is a function of $I(t)$, that is, $f_{m}(I(t))$, as the source of media depends on the density of symptomatic populations. A schematic representation of the above biological mechanism of coronavirus in human upon which our model is based is shown in the Figure \ref{Schema}. The dynamics of novel coronavirus model is governed by the following system of nonlinear ordinary differential equations:

\subsection{Dynamics of Susceptible individuals $S(t)$}

\noindent The susceptible individual is recruited into the population with a constant inflow rate $\Lambda_{s}$ and decreased by a natural mortality mortality rate  $\delta_{s}$.
Here, $\beta_{s}\alpha_{a}$ represents the transmission coefficient of the asymptomatic individuals to susceptible, $\beta_{s}\alpha_{i}$ is the transmission coefficient of symptomatic infected classes to the susceptible, and $\beta_{s}\alpha_{h}$ denotes transmission coefficient of the hospitalized classes to susceptible population. We consider $\beta_s$ is the disease transmission coefficient for both the infectiousness of COVID-19 and contact rates, with adjustment factors $\alpha_a$ (for asymptomatic classes),  $\alpha_i$ (for symptomatic classes)  and $\alpha_h$ (for hospitalized individuals).

Albeit, coronavirus is supposed to be spread completely by symptomatic populations, a very lower rate of transmission by asymptomatic populations cannot even be ruled out. The adjustment parameter $\alpha_a \geq 0$ take into accounts for varying levels of hygiene preventions through asymptomatic, an analogous explanation can be drawn to the adjustment parameters $\alpha_i $  and $\alpha_h $ during symptomatic and hospitalization, respectively. As the asymptomatic, symptomatic and hospitalization programmes and hygiene preventions during asymptomatic, symptomatic and hospitalization were executed and enriched progressively after an epidemic of coronavirus, the spreading coefficients $\alpha_a \beta_s $, $\alpha_i\beta_s $ and $\alpha_h\beta_s$, asymptomatic, symptomatic and hospitalized rates could be modeled as a time-dependent parameters in computations. Moreover, the interplay among susceptible and infected population (asymptomatic, symptomatic and hospitalized) is modeled utilizing standard homogeneous mixing incidence \cite{Anderson91,Diekmann00}, in the form of total individuals. The rate of change of susceptible individuals can be represented by the following ordinary differential equation (ODE):

\begin{eqnarray}
\label{stateeq1}
\frac{dS}{dt} &=& \Lambda_{s} - \frac{\beta_{s} S (\alpha_{a}A + \alpha_{i}I + \alpha_{h}H)}{N} - \delta_{s}S - f(S, M).
\end{eqnarray}

\subsection{Dynamics of Exposed individuals $E(t)$}
\noindent The population who are exposed individuals but not yet developed clinical symptoms of coronavirus. The exposed individual is declined by asymptomatic individuals at the rate $\gamma_e$ and natural death at the rate $\delta_e$. The constant per-capita rate $\gamma_e$ modeled progression from the exposed individuals $(E)$ to either the asymptomatic individuals $(A)$ or symptomatic individuals $(I)$; the constant $q_a (0 < q_a < 1)$ represents budget for the rate of progress of population who move to either the asymptomatic partially-infectious class $(A)$ or  symptomatic class $(I)$ at the per-capita rate $\gamma_e$. The corresponding model dynamics for exposed population can be presented by the following ODE:

\begin{eqnarray}
\label{stateeq2}
\frac{dE}{dt} &=& \frac{\beta_{s} S (\alpha_{a}A + \alpha_{i}I + \alpha_{h}H)}{N} - \delta_{e} E - \gamma_{e} E.
\end{eqnarray}

\subsection{Dynamics of Asymptomatic individuals $A(t)$}

\noindent Asymptomatic populations were exposed to the coronavirus but not infectious for the community. The exposed classes move to the asymptomatic classes at the rate $\gamma_{e}$ by a constant portion $(1 - q_a)$. The asymptotic individuals progress to the recovered class at the per-capita rate $\xi_a$ and natural mortality rate $\delta_a$. The rate of change of asymptomatic individuals is govern by the following ODE:

\begin{eqnarray}
\label{stateeq3}
\frac{dA}{dt} &=& (1 - q_{a}) \gamma_{e} E - \delta_{a} A - \xi_{a} A.
\end{eqnarray}

\subsection{Dynamics of Symptomatic individuals $I(t)$}

\noindent The symptomatic individuals are produced after the progression of clinical symptoms of coronavirus (diagnosed by the clinicians) by the asymptomatic individuals.
The exposed individuals move to the symptomatic classes at the per-capita rate $\gamma_{e}$ by a constant portion $q_a$. The symptomatic or clinically ill individuals progress to the hospitalized or isolated class $(H)$ at the per-capita rate $\gamma_i$ and natural mortality rate $\delta_i$. It can be assumed that $\gamma_i$, the per-capita rate at which symptomatic or clinically ill classes seek clinical attention and are therefore put into the hospitalized class, is bigger than $\gamma_e$, the per-capita rate at which the exposed individuals progress to the asymptomatic classes. The clinically ill classes (class $I$) are moved to the recover classes without being diagnosed at the per-capita rate $\xi_i$. The transmission dynamics of symptomatic populations are governed by the following ODE:

\begin{eqnarray}
\label{stateeq4}
\frac{dI}{dt} &=& q_{a} \gamma_{e} E - \delta_{i} I - \xi_{i} I - \gamma_{i}I.
\end{eqnarray}

\subsection{Dynamics of Hospitalized individuals $H(t)$}

\noindent These are the populations who have progressed clinical symptoms for coronavirus and have been isolated at the hospital for medical treatment. The hospitalized individuals are came from the clinically ill classes or symptomatic community (class $I$) at the per-capita rate $\gamma_i$ and it has natural mortality rate $\delta_h$. Hospitalized or isolated populations recover at the per-capita rate $\xi_h$. We assumed that $\xi_h > \xi_i$, as an hospitalized populations are likely to obtain a partially efficacious during isolation or hospitalization. The dynamics of hospitalized individuals are modeled by the following ODE:

\begin{eqnarray}
\label{stateeq5}
\frac{dH}{dt} &=& \gamma_{i} I - \delta_{h} H - \xi_{h} H.
\end{eqnarray}

\subsection{Dynamics of Recovered individuals $R(t)$}

\noindent It can be assumed that the recovered population have enduring immunity against coronavirus albeit to date this has neither been confirmed nor contradicted. The asymptomatic, symptomatic and hospitalized individuals recover from the coronavirus at the per-capita rates $\xi_a$, $\xi_i$ and $\xi_h$, respectively. The recovered individuals die at the per-capita rate $\delta_r$. The rate of change of recovered individuals can be written by the ODE:

\begin{eqnarray}
\label{stateeq6}
\frac{dR}{dt} &=& \xi_{a}A + \xi_{i}I + \xi_{h}H - \delta_{r} R + f(S, M).
\end{eqnarray}

\subsection{Information variable or media awareness program $M(t)$}

\noindent The function $f(S, M)$ indicates the behavioral response for the susceptible population, induced by the effects of media related information of the prevalence of coronavirus, by considering accessible precautionary measures to become aware about the coronavirus disease. The behavioral response is not completely efficacious due to the financial assistance, lack of the proper resources and an irresponsive behavior among the populations, we consider the media term as $f(S, M) = \eta_m d_m S M$. Here, $\eta_m$ represents the corresponding response rate and $0 \leq \eta_m \leq 1$ determines the response intensity. The parameter $d_m$ indicates the media related information index rate by which the population alter their behavior due to the prevalence of coronavirus. The parameter $\delta_m$ represents the natural degradation of media related information index, which can be occurred due to the fading of memory about media related information with time in addition to the complacent behavior of population.  \\

We assume that the source of media related information index is proportional to the number of symptomatic individuals. The function $f_m(I)$ represents the growth of information index rate, which depend on the symptomatic individuals. While the coronavirus outbreaks, the World Health Organization (WHO), health agencies, social media, government and non-government organization become dynamic to spread the information or media awareness in concerning the prevalence of coronavirus.  \\

As for the function $f_m(I)$, one can consider any function satisfying $f_m(0) = 0$ and $f'_m(I) > 0$ \cite{Buonomo08}. The most easiest example is $f_m(I) = \alpha_m I$ with $\alpha_m \in (0, 1)$, that is, $I$ is linear function of the present prevalence of the coronavirus, which could represent for instance current standardized incidence of serious cases of the coronavirus \cite{d'Onofrio07}. One more example could be $f_m(I) = \frac{\alpha_m I}{1 + \beta_{m} I}$, where $I$ indicates the nonlinear increasing function for the prevalence of the coronavirus, which could be considered as a measure of the perceived risk for the infection rather than the original one \cite{d'Onofrio09, Reluga06}. Here $\alpha_m$ represents the source rate of the information index and $\beta_{m}$ is half saturation constant. Thus, we consider the following ODE model for media awareness program:

\begin{eqnarray}
\label{stateeq7}
\frac{dM}{dt} & =& f_{m}(I) - \delta_{m}M.
\end{eqnarray}

Based on the above biological assumptions and schematic representation of coronavirus (see the Figure \ref{Schema}) together with specific forms, we formulate a mathematical model of novel coronavirus with media awareness given by the following seven-dimensional nonlinear system of ODEs:

\begin{eqnarray}
\label{stateeq}
\left\{
  \begin{array}{ll}
    \frac{dS}{dt} = \Lambda_{s} - \frac{\beta_{s} S (\alpha_{a}A + \alpha_{i}I + \alpha_{h}H)}{N} - \delta_{s}S - \eta_m d_m S M, \\
   \frac{dE}{dt} = \frac{\beta_{s} S (\alpha_{a}A + \alpha_{i}I + \alpha_{h}H)}{N} - \delta_{e} E - \gamma_{e} E, \\
   \frac{dA}{dt} = (1 - q_{a}) \gamma_{e} E - \delta_{a} A - \xi_{a} A,  \\
    \frac{dI}{dt} = q_{a} \gamma_{e} E - \delta_{i} I - \xi_{i} I - \gamma_{i}I, \\
    \frac{dH}{dt} = \gamma_{i} I - \delta_{h} H - \xi_{h} H, \\
    \frac{dR}{dt} = \xi_{a}A + \xi_{i}I + \xi_{h}H - \delta_{r} R + \eta_m d_m S M, \\
    \frac{dM}{dt} = \alpha_{m}I - \delta_{m}M,
  \end{array}
\right.
\end{eqnarray}
the model is satisfied the following positive initial conditions:
\begin{eqnarray}
\label{IC}
S(0) = S_0,~~~E(0)=E_0,~~~A(0)=A_0,~~~I(0)=I_0,~~~H(0)=H_0,~~~R(0)=R_0,~~~M(0)=M_0.
\end{eqnarray}
Since the model system (\ref{stateeq}) monitors the dynamics of human population, all state variables are assumed to be positive.

\subsection{Basic reproduction number}

\noindent The basic reproduction number, symbolized by $R_0$, is `the expected number of secondary cases produced, in a completely susceptible population, by a typical infective individual' \cite{Diekmann90,PVDD02}. The dimensionless basic reproduction number provides a threshold, which play a crucial role in determining the disease persists or dies out from the individual. In a more general way $R_0$ can be stated as the number of new infections created by a typical infective population at an infection free equilibrium. $R_0 < 1$ determines on average an infected population creates less than one new infected population during the course of its infective period, and the infection can die out. In reverse way, $R_0 > 1$ determines each infected population creates, on average, more than one new infection, and the disease can spread over the population. The basic reproduction number $R_0$ can be computed by using the concept of next generation matrix \cite{Diekmann90,PVDD02}. In order to do this, we consider the nonnegative matrix $\mathcal{F}$ and the non-singular $M-$matrix $\mathcal{V}$, expressing as the production of new-infection and transition part respectively, for the system (\ref{stateeq}), are described by
\begin{eqnarray*}
\mathcal{F} &=& \left[ \begin{array}{c}
       \frac{S}{N} \beta_s (\alpha_a A + \alpha_i I + \alpha_h H)  \\
       0 \\
       0 \\
       0
     \end{array} \right],~~~~~~~~~\mathcal{V} = \left[ \begin{array}{c}
       (\delta_e + \gamma_e) E  \\
       - (1 - q_a)\gamma_e E + (\delta_a + \xi_a) A \\
       - q_a \gamma_e E + (\delta_i + \xi_i + \gamma_i) I \\
       -\gamma_i I + (\delta_h + \xi_h) H
     \end{array} \right].
\end{eqnarray*}

The Jacobian matrix for the system (\ref{stateeq}), can be computed at an infection free state $(E = A = I = H = 0)$, we have
\begin{eqnarray*}
F &=& \left[ \begin{array}{cccc}
       0 & \beta_s \alpha_a & \beta_s \alpha_i & \beta_s \alpha_h \\
       0 & 0 & 0 & 0 \\
       0 & 0 & 0 & 0 \\
       0 & 0 & 0 & 0
     \end{array} \right]
~~~~~~~~V ~=~ \left[ \begin{array}{cccc}
       \delta_e + \gamma_e & 0 & 0 & 0 \\
       - (1 - q_a)\gamma_e & (\delta_a + \xi_a) & 0 & 0 \\
       - q_a \gamma_e & 0 & (\delta_i + \xi_i + \gamma_i) & 0 \\
       0 & 0 & -\gamma_i & (\delta_h + \xi_h)
     \end{array} \right]
\end{eqnarray*}

The basic reproduction number $R_0 = \rho(FV^{-1})$, where $\rho(FV^{-1})$ represents the spectral radius for a next generation matrix $FV^{-1}$. Thus, from the system (\ref{stateeq}), we get the basic reproduction number $R_0$ is
\begin{eqnarray}
\label{R0}
\rho(F V^{-1}) &=& R_{0} = \frac{(1 - q_a)\gamma_e \beta_s \alpha_a}{(\delta_e + \gamma_e) (\delta_a + \xi_a)} + \frac{q_a \gamma_e \beta_s \alpha_i}{(\delta_e + \gamma_e) (\delta_i + \xi_i + \gamma_i)} + \frac{q_a \beta_s \alpha_h \gamma_e  \gamma_i}{(\delta_e + \gamma_e) (\delta_i + \xi_i + \gamma_i) (\delta_h + \xi_h)}.~~~~~~~~~
\end{eqnarray}

\section{Model calibration and coronavirus data source}

\noindent A detailed elucidation on the computer simulations for the coronavirus model system (\ref{stateeq}) is investigated in this section. We start by building up on the parameter estimations followed by some computer simulation results, which provide insights on the eradication and development of the dynamics of novel coronavirus (COVID-19) epidemic. Parameter values are taken from literature and some of the parameters are estimated from the observed data. List of the parameter values with description are given in the Table \ref{parval} and the values of the remaining parameters estimated from the observed data, which are given in the Table \ref{parvalEst}. Due to the lack of the data of coronavirus diseases related to the effect of media related awareness, we have simulated our model without the effect of media awareness. But, we have simulated our model with the effect of media related awareness for four different states, namely Delhi, Maharashtra, Kerala and West Bengal as well as the Republic of India.
 The initial population sizes for different states, namely Kerala, Maharashtra, Delhi, West Bengal and the Republic of India are stated in the Table \ref{parvalIC}. Also the recruitment rate $\Lambda_s$ for four different states and overall India are listed in the Table \ref{parvalIC}.

\subsection{PRCC sensitivity analysis}

\noindent To identify the most sensitive parameters with respect to symptomatic population, we performed a sensitivity analysis by using Partial Rank Correlation Coefficient (PRCC) technique for all the input parameters against the variable $I$. The set of most sensitive parameters, which have the ability to reduce the outbreak of coronavirus diseases, can be estimated from the available data from different states (Kerala, Maharashtra, Delhi, West Bengal) as well as the scenario of entire India. In our coronavirus system (\ref{stateeq}) has 17 parameters (without media) for which we varied all the system parameters simultaneously. PRCC quantifies the relationship among a state equation of interest and each of model parameters. As a result PRCC aid to obtain the important parameters, which contribute most to the system variability. For model simulation, we consider the PRCC values between -1.0 and +1.0.  \\

\begin{figure}
  \centering
  \includegraphics[width=1.0\textwidth]{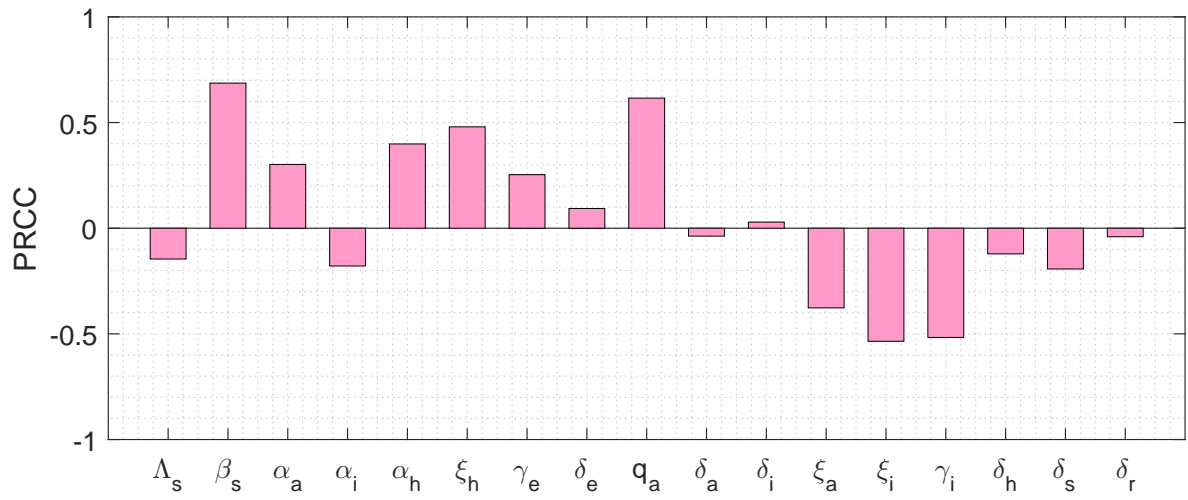} \\
  \caption{ Partial rank correlation coefficients illustrating the dependence of symptomatic individuals $I$ on each of the system parameters at the day 60 with $p < 0.02$.}
\label{Prcc}
\end{figure}

Following the method established by Marino et al. \cite{Marino08}, we performed Latin hypercube sampling and generated 2800 samples to compute the PRCC and the p-values with reference to the symptomatic classes at the day 60. The Figure \ref{Prcc} designates the PRCC results, which implies that the highest positively correlated parameters are the disease transmission coefficient rate $\beta_s$, proportion rate of exposed individuals $q_a$, the rate $\xi_h$ by which the hospitalized class converted to recovered class, the rate $\alpha_h$ by which susceptible individuals converted to exposed class and the highly negatively correlated parameters are the rate $\xi_i$ by which symptomatic individuals become recovered, the rate $\gamma_i$ by which symptomatic individuals converted to hospitalized class, and the rate $\xi_a$ by which the asymptomatic individuals moved to recovered individuals, account for most uncertainty with respect to the symptomatic population. Thus, our PRCC analysis yields these 7 parameters $\beta_s$, $q_a$, $\xi_i$, $\gamma_i$, $\xi_h$, $\alpha_h$ and $\xi_a$ are the most effective parameters out of 17 parameters. Therefore, we estimated the 7 parameters by using least square method.

\subsection{Parameter estimation}

\noindent Now, we have calibrated our coronavirus disease model (\ref{stateeq}) with daily confirmed new coronavirus cases for the entire India and four states namely, Kerala, Delhi, West Bengal and Maharashtra. Data of daily confirmed new coronavirus cases for whole India obtained from World Health Organization (WHO) situation report for the period January 30, 2020 - March 30, 2020. Data of daily confirmed new coronavirus cases for the state West Bengal obtained from Health \& Family Welfare Department, Government of West Bengal for the period March 17, 2020 - March 31, 2020. Data of daily confirmed new coronavirus cases for the state Kerala obtained from the Directorate of Health Services, Government of Kerala for the period January 30, 2020 - March 31, 2020. The data of daily confirmed new coronavirus cases for the state Delhi obtained from the Health \& Family Welfare, Govt. of the National Capital Territory of Delhi for the period March 4, 2020 - March 30, 2020. Also, the data of total coronavirus cases for the state Maharashtra obtained from India COVID-19 Tracker (https://www.covid19india.org/) for the period March 15, 2020 - March 31, 2020. We fit our coronavirus model system (\ref{stateeq}) (without media related awareness program) with daily new coronavirus cases for the four states (Delhi, Kerala, Maharashtra and West Bengal) and for India. Also, the cumulative coronavirus cases has been fitted with our model system. The observed data were fitted in the optimization toolbox of Matlab using nonlinear least squares method and we have estimated the parameters $\beta_s$, $\xi_a $, $q_a$, $\xi_h$, $\gamma_i$, $\xi_i$ and $\alpha_h$ as these parameters found more sensitive in the PRCC analysis. The biological interpretations of the model parameters are given in the Table \ref{parval}. The estimated values of the model parameters for all the data sets are given in the Table \ref{parvalEst}. The initial population sizes for all the data sets are also estimated to match with the confirmed new coronavirus cases (see Table \ref{parvalIC}). The output of our model for daily new cases are shown in Figure \ref{DailyNewCases}. Also, the observed daily new coronavirus cases are overlapped on the Fig. \ref{DailyNewCases}. It is also important to observe the total number of coronavirus cases in a province only. So, the observed daily cumulative confirmed coronavirus cases and the model simulation has been plotted in the Fig. \ref{CumulativeCases}.    \\

To assure the performance of our coronavirus system without media (\ref{stateeq}), we have computed Mean Absolute Error $(E_{MAE})$ and Root Mean Square Error $(E_{RMSE})$. Here, $E_{MAE}$ and $E_{RMSE}$ are given by

\begin{eqnarray*}
E_{MAE} &=& \frac{\Sigma_{i=1}^{n} \vert O(i)-Q(i) \vert}{n},  \\
E_{RMSE} &=& \sqrt{\frac{\Sigma_{i=1}^{n} ( O(i)-Q(i))^2}{n}},
\end{eqnarray*}

where $O(i)$ represents the observed value, $Q(i)$ is the model output and $n$ is the sample size of the observed data. The values of $E_{MAE}$ and $E_{RMSE}$ for the four states (Delhi, Kerala, Maharashtra and West Bengal)  and the Republic of India are presented in Table \ref{error}.   \\

The fitting data enables us to quantify the basic reproduction number $R_0$ from the expression of (\ref{R0}). The basic reproduction number for coronavirus in different states, for Kerala, $R_0 = 4.78435$; Delhi, $R_0 = 3.28271$; West Bengal, $R_0 = 2.73854$; Maharashtra, $R_0 = 3.03971$; and for India, $R_0 = 2.58792$. Let us observe that the basic reproduction is relatively high in Kerala, relative to the other three states and the overall India considered here. From the basic reproduction number $R_0$ in overall India, it can be conclude that the coronavirus outbreaks is pandemic and the Govt. should take a necessary action to keep social distancing and necessary precautions. The basic reproduction number $R_0$ for four provinces and overall India are  presented the Table \ref{parvalR0}.  \\

From the data fitting, it can be observed that our model performs relatively good in all the states namely, Kerala, Delhi, West Bengal, Maharashtra. This is showing a good agreement with the observed data and curve fitting with the coronavirus disease model (\ref{stateeq}) under consideration.  However, the model performance less well in the case of whole India as the number of confirmed coronavirus cases is relatively high. Though, the increasing trend of the daily confirmed coronavirus cases have been nicely captured by the model for all the data sets. We also compute the numerical errors for computing the numerical simulations for our model and the corresponding  lowest value of $E_{MAE}$ and $E_{RMSE}$ have been obtained for the state West Bengal, since the size of the data is lower than the other data sets.

\subsection{Short-term prediction}

\noindent Due to absence of any effective vaccine or treatment strategies for novel coronavirus,  predictive mathematical models can aid in understanding the complicated dynamics of coronavirus transmission and its control as well as its intricate epidemiological cycle. In order to study the short term prediction for our model system (\ref{stateeq}), we have simulated the cumulative symptomatic classes of coronavirus cases for the time period April 01, 2020 to April 10, 2020 using the estimated parameters values in the Table \ref{parvalEst} with initial population sizes in the Table \ref{parvalIC}. The other baseline parameter values are specified in the Table \ref{parval}. The simulated results for the period April 01, 2020 to April 10, 2020 for the three provinces, namely  Kerala, West Bengal, Maharashtra and the union territory Delhi and the Republic of India have shown in the Figure \ref{f_pred}. The black solid curves are the model simulation and the solid black circles are the observed values, as plotted in the Figure \ref{f_pred}. Our model simulation represents that the model performs well and successfully able to capture the increasing trends of all the four cases (Delhi, Maharashtra, West Bengal and the overall India) whereas the slope of the simulated curve is relatively high for the state Kerala.

\subsection{Sensitivity analysis for $R_0$}

\noindent To describe how best to diminish human impermanency and morbidity due to novel coronavirus, it is very essential to see the relative significance of various factors responsible for its transmission. The initial disease transmission is completely associated to the reproduction number $R_0$, and we calculate the sensitivity indices with regard to $R_0$, to the parameters for the system (\ref{stateeq}). The sensitivity indices describe us whether or not the infectious diseases will develop throughout the individulas. Sensitivity analysis is mainly utilized to describe the robustness of the reproduction number $R_0$ to fluctuation in the system parameters. It also give us to identify the relative change in a state variable when a system parameter alters. The normalized forward sensitivity index for $R_0$ with respect to the parameter $p$ is defined as follows:
\begin{eqnarray*}
\Gamma^{R_0}_{p} &=& \frac{\partial R_0}{\partial p} \times \frac{p}{R_0}.
\end{eqnarray*}
As we have the explicit expression for $R_0$ in (\ref{R0}), we obtain the analytical formulae for sensitivity indices of $R_0$, $\Gamma^{R_0}_{p}$, to each of fourteen parameters stated in the Table \ref{R0_sens}.

As for example, the sensitivity index of $R_0$ with regards to $\beta_s$ is
\begin{eqnarray*}
\Gamma^{R_0}_{\beta_s} &=& \frac{\partial R_0}{\partial \beta_s} \times \frac{\beta_s}{R_0} ~=~ 1,
\end{eqnarray*}
independent of any parameters. This represents that $R_0$ is the increasing function with respect to $\beta_s$. It implies that the probability of disease transmission has a high impact on coronavirus control and management. Other parameters have their sensitivity indices are specified in the Table \ref{R0_sens}.

From the Table \ref{R0_sens}, it can be noticed that some of the indices are positive and some of the indices are negative, which indicate that the indices having `+' signs increase the value of $R_0$  as one increase them and those having `-' signs decrease the value of $R_0$, as they are increased. The most sensitive parameter is the disease transmission coefficient $\beta_s$, where sensitivity index $\Gamma^{R_0}_{\beta_s} ~=~ 1$ indicates that the increasing $\beta_s$ by 10\% will increase $R_0$ by 10\%. Also, the sensitivity index of $\xi_a$, that is, $\Gamma^{R_0}_{\beta_s} ~=~ - 0.844286$ represents that increasing $\xi_a$ by 10\% will decrease $R_0$ by 8.44286\%. Similarly, we can interpret the other parameters too.

We plot the sensitivity indices of $R_0$, for each of the model parameters using in the Table \ref{R0_sens} has been shown in the Figure \ref{Fsens_r0}. In order to control the coronavirus outbreaks, we must target the most sensitive parameters who has the great impact in controlling the system dynamics. For instance, the disease transmission coefficient $\beta_s$ is the most effective parameter in controlling the coronavirus diseases, which can easily be identified from the Figure \ref{Fsens_r0} as well as in the Table \ref{R0_sens}.

To control the coronavirus outbreaks, the draw the contour plots for the $R_0$ with respect to the positively correlated parameters, namely disease transmission coefficient $\beta_s$ and the clinical outbreak rate $q_a$ for all the infected classes in the Figure \ref{contour}. The contour plot represents the dependence of $R_0$ for the four states (Kerala, Delhi, West Bengal, Maharashtra) and the Republic of India. Contour plot indicates that for the higher values of $\beta_s$ and $q_a$ the reproduction number $R_0$ become high, which means the coronavirus may enter the stage-3 and it spread through the community. Therefore, to control $R_0$ must be restrict the values of the transmission coefficient $\beta_s$ and the portion $q_a$ by which the exposed class become symptomatically ill. To keep the mortality rate as low as possible, we must be take the values of $\beta_s$ and $q_a$ as small as possible. The Figure \ref{Fsens_r0} and Table \ref{R0_sens} represents that the adjust factors $\alpha_a$, $\alpha_i$ and $\alpha_h$ are positively correlated with the basic reproduction number $R_0$. Thus, we must have to reduce the adjust factors too. Thus, we may conclude that to reduce the coronavirus disease outbreak we need to maintain social distancing, limit or stop taking fairs and theaters performances etc.

\subsection{Effect of media related awareness}

\begin{figure}
\centering
\includegraphics[width=15cm]{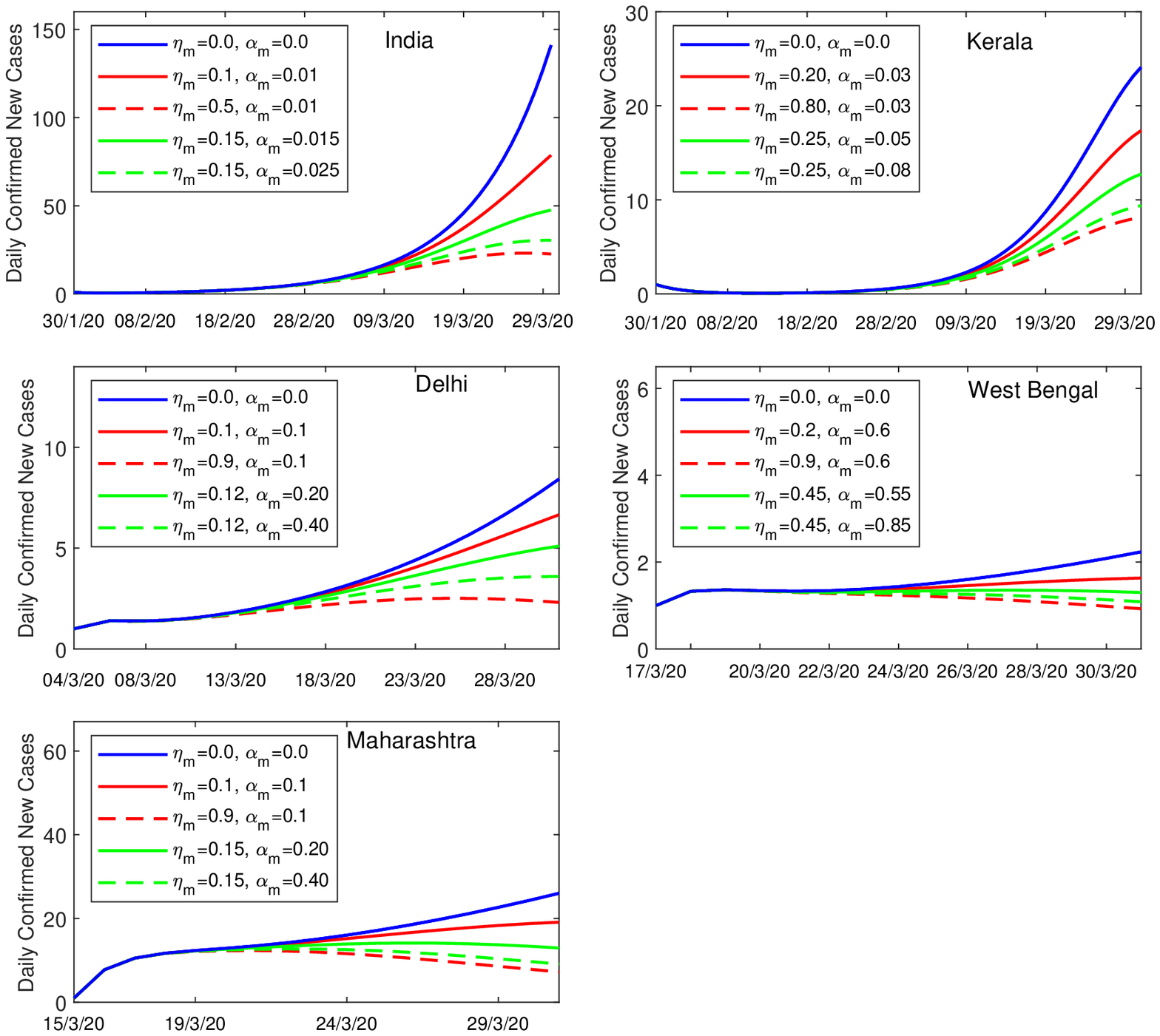}
\caption{Illustrations of the media related effect for the daily confirmed new cases of coronavirus infected patients for four different states (Kerala, Delhi, Maharashtra, West Bengal) and the Republic of India. The model baseline parameters are taken from the Table \ref{parval} and rest of parameters are estimated and taken from the Table \ref{parvalEst} for different states and overall India. The initial values are taken from the Table \ref{parvalIC}. }
\label{f-media-new}
\end{figure}

\noindent To study the impact of media related awareness for the coronavirus disease we simulate our system (\ref{stateeq}) for four provinces, namely Kerala, Maharashtra, Delhi, West Bengal and the Republic of India, with respect to the parameter values specified in the Table \ref{parval} and rest of the parameter values estimated and are given in the Table \ref{parvalEst} with initial population sizes are given in the Table \ref{parvalIC}. The Figure \ref{f-media-new} represents the simulation of our model with media awareness for daily confirmed new cases of coronavirus of infected patients. The first sub-figure of the Figure \ref{f-media-new} represents the corresponding profiles of symptomatic individuals for the overall India in absence and presence of media awareness programme with the available time period 60 days. Observe that the blue curve has the highest peak or the proliferation of symptomatic population is high and sharp in compare to the other curves, which indicates that the sizes of symptomatic individuals increases and reaches to its maximum peak in absence of media awareness as the source of information index $(\alpha_m)$ and the rate of media response $(\eta_m)$ both are zero. To investigate how often and/or how long the media awareness effect remains dynamic, we simulate our coronavirus disease model (\ref{stateeq}). But in presence of active media awareness, both $\alpha_m$ and $\eta_m$ are nonzero, and for fixed $\alpha_m = 0.01$ when we varied the rate of media response $\eta_m = 0.1$ and $\eta_m = 0.5$, the thick red curve and dotted red curve decreases but the peak of both the red curves are below the thick blue curve. For fixed media response $\eta_m = 0.15$ when we varied the source rate of information index $\alpha_m = 0.015$ and $\alpha_m = 0.025$, it can be noted that the thick green curve and dotted green curve decreases but the peak of both the green curves are below the thick blue cure. The model simulation (in case of overall India) showed in the first figure of Figure  \ref{f-media-new}, it can be noted that in presence of media effect the count of symptomatic individuals decreases. It is worthy to mention that the media has an impact on the prevalence of the outbreak of coronavirus diseases. It is fascinating to observe that the media effect remains dynamic almost until the highest peak of the disease outbreak, has been shown in the Figure \ref{f-media-new}. Similar interpretations can be made for the four provinces of the Republic of India, namely Delhi, Maharashtra, Kerala and West Bengal, in presence of media related awareness programme for daily confirmed new cases of coronavirus diseases of symptomatic individuals. From the model simulation, we can conclude that the media awareness has an impact on the reduction of coronavirus diseases for the symptomatic individuals, as shown in the Figure \ref{f-media-new}.

\begin{figure}
\centering
\includegraphics[width=15cm]{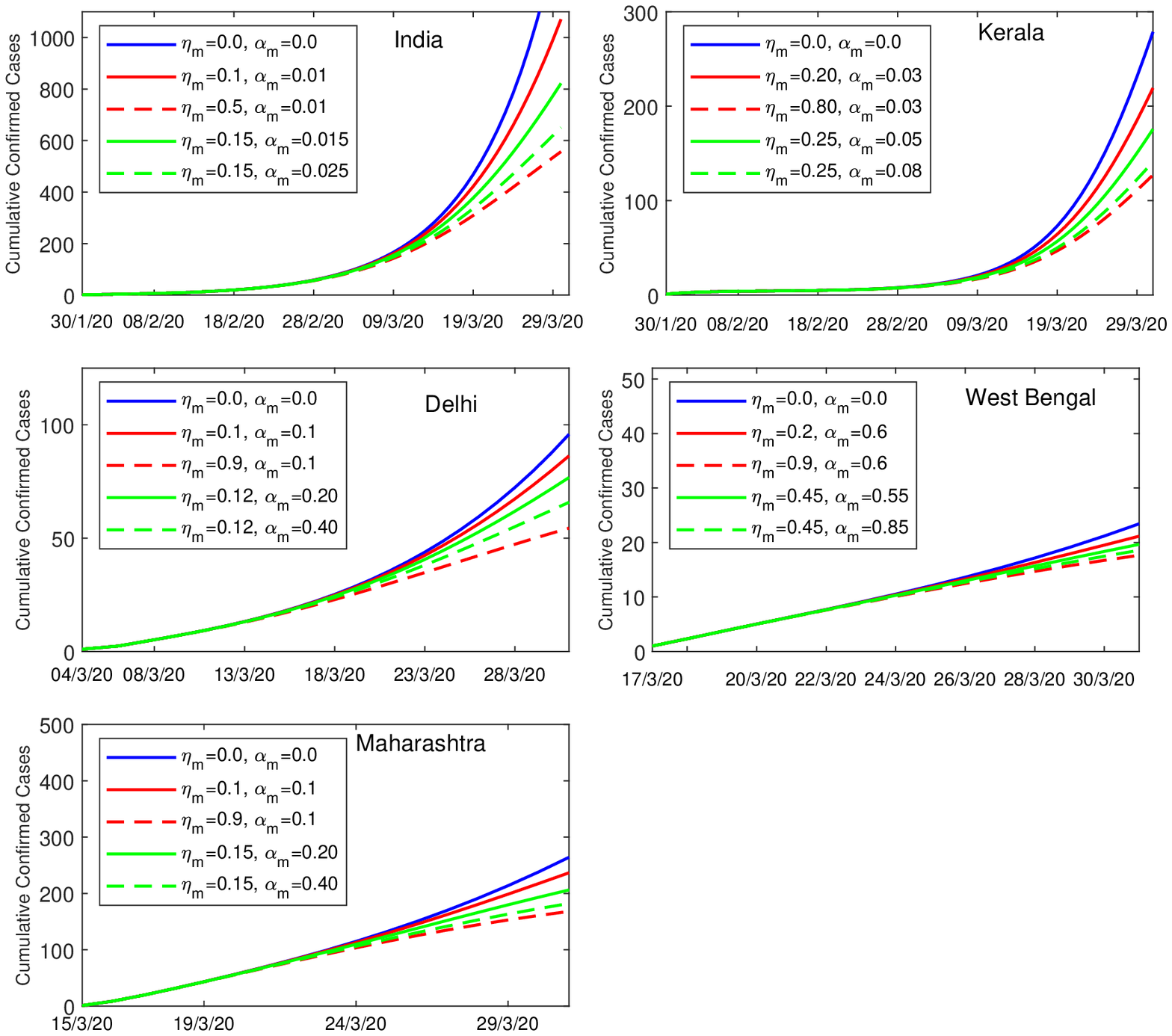}
\caption{Illustrations of the media related effect for the daily cumulative confirmed cases of coronavirus infected patients for four different states (Kerala, Delhi, Maharashtra, West Bengal) and the Republic of India. The model baseline parameters are taken from the Table \ref{parval} and rest of parameters are estimated and taken from the Table \ref{parvalEst} for different states and overall India. The initial values are taken from the Table \ref{parvalIC}.}
\label{f-media-cumul}
\end{figure}

We also investigate the media related awareness for the novel coronavirus diseases for the Republic of India and four provinces, namely Kerala, Maharashtra, West Bengal and Delhi, with respect to parameters listed in the Table \ref{parval} and rest of the estimated parameters listed in the Table \ref{parvalEst} with initial individuals are listed in the Table \ref{parvalIC}. The impact of media awareness effect for the daily cumulative confirmed cases of coronavirus diseases for symptomatic classes are shown in the Figure \ref{f-media-cumul}. In the first sub-figure of the Figure \ref{f-media-cumul} indicates the model simulation of (\ref{stateeq}) corresponding to the profiles of symptomatic individuals for overall Indian scenarios with the available time period 60 days. In this case the thick blue curve showing the maximum peak or the growth of cumulative symptomatic individuals in compare to other curves, which represents that the sizes of cumulative symptomatic or clinically ill population reached their maximum peak in absence of media information as the source of information index $(\alpha_m)$ and the rate of media response $(\eta_m)$ both are considered as zero. In presence of effective media awareness, that is, both $\alpha_m$ and $\eta_m$ are nonzero, and for fixed $\alpha_m = 0.01$ when varied the media response rate $\eta_m = 0.1$ and $\eta_m = 0.5$ the thick red curve and dotted red curve decreases significantly and both the peak remains below the thick blue curve. The same scenarios has been observed when we vary information index rate $\alpha_m = 0.015$ and $\alpha_m = 0.025$ with fixed value for media response rate $\eta_m = 0.15$ the thick green curve and dotted blue curve reduces noteworthy and remains below the thick blue curve. Thus, the media related awareness reduces the peak of symptomatic individuals in case of cumulative confirmed coronavirus diseases, which shows that the media awareness play a key role on the prevalence of the outbreak of coronavirus diseases, as shown in the Figure \ref{f-media-cumul}. It is very interesting to note that the media effect remains effectual almost until the maximum peak of the disease epidemic. Similar conclusions can be drawn for four provinces of India, namely Kerala, Maharashtra, Delhi and West Bengal in presence of effective media awareness for daily cumulative confirmed cases of symptomatic populations. Thus, the media awareness has a great impact to control the coronavirus diseases, as shown in the Figure \ref{f-media-cumul}.

\subsection{Long-term dynamics}

\noindent To study the future outbreaks of coronavirus diseases, we simulate our system (\ref{stateeq}) for a long days of the infected individuals, namely asymptomatic individuals and symptomatic or clinically ill individuals for the parameters are listed in the Table \ref{parval} and we have considered the estimated parameters for the Table \ref{parvalEst} in case of Indian data only with initial population sizes are given in the Table \ref{parvalIC}. The model simulation represented that the people from overall India can experience future epidemics if the control policies are not implemented more effectively, as shown in the Figure \ref{f-longTime}. The number of infected individuals (asymptomatic and symptomatic) are oscillate around 130 days and around 320 days. This can be interpreted as the reproduction number $R_0$ is always greater than unity in India (see the Table \ref{parvalR0} for India). Otherwise, we can say that the impact of media related awareness is a dynamic process and the media has a great impact in reducing the transmission of coronavirus outbreaks at the initial stage of an epidemic. The factors like low air temperature and as well low humidity has a great impact on the transmission of novel coronavirus \cite{Jingyuan20}. Moreover, to control the future outbreaks of novel coronavirus the media related awareness programme and the control strategy must be implemented properly.

\section{Discussion}

\noindent According to the INDIA COVID-19 TRACKER report, 7,598 confirmed cases and 246 deaths due to novel coronavirus outbreaks in India \cite{Indiacov19}, as of April 10, 2020. The total number of new cases and new deaths are occurred everyday reported from different cities across the India \cite{Indiacov19}. This is a terrifying circumstances as India has approximately 139 crores population and may enter stage-3 of coronavirus disease transmission. Till date, there is no specific vaccine, antivirals or effective therapeutics to treat coronavirus diseases. During this period, forecasting is the utmost priority for the planning of heath care and control the novel coronavirus diseases. Thus, mathematical modeling can aid us in designing to control the outbreak of coronavirus diseases in absence of any treatment or any specific diagnostic test.

In this manuscript, we propose a compartmental epidemic model on novel coronavirus diseases by incorporating the effect of media awareness to predict and control the outbreak. In our model we noted that media awareness play a key role in generating public consciousness and encouraging disease measures \cite{Funk10}. In our proposed model, we consider two infectious individuals, namely asymptomatic and symptomatic individuals with former being a fast spreader of the coronavirus diseases. Our model formulation is based on the work by Gumel et al.  \cite{Gumel04}, where we estimate the system parameters based on the PRCC sensitivity analysis at an early stage of the outbreak, in which the number of cumulative classes grow exponentially.

We calibrated the proposed model to fit with the data from four different states, namely Maharashtra, Kerala, Delhi, West Bengal and the Republic of India. The most important part is to perform the sensitivity analysis of 17 model parameters and its estimation. To identify the key input parameters which contributed to the symptomatic outcome, strongly assisted the implementation of an integrated policy for various measures, involving the epidemic during different phases of the diseases, we use the sensitivity analysis by using PRCC techniques. PRCC analysis yields the most sensitive parameters are $\beta_s$, $q_a$, $\gamma_i$, $\alpha_h$, $\xi_a$ and $\xi_h$ out of 17 input parameters without incorporating media impact. Then we estimated these 7 parameters based on the data from four different states and from the entire India. By fitting data on novel coronavirus cases during January 30, 2020 to March 31, 2020, in the four different states of India and the overall India to our proposed model, we were able to get estimates of the unknown parameters for our system.

\begin{figure}
\centering
\includegraphics[width=15cm]{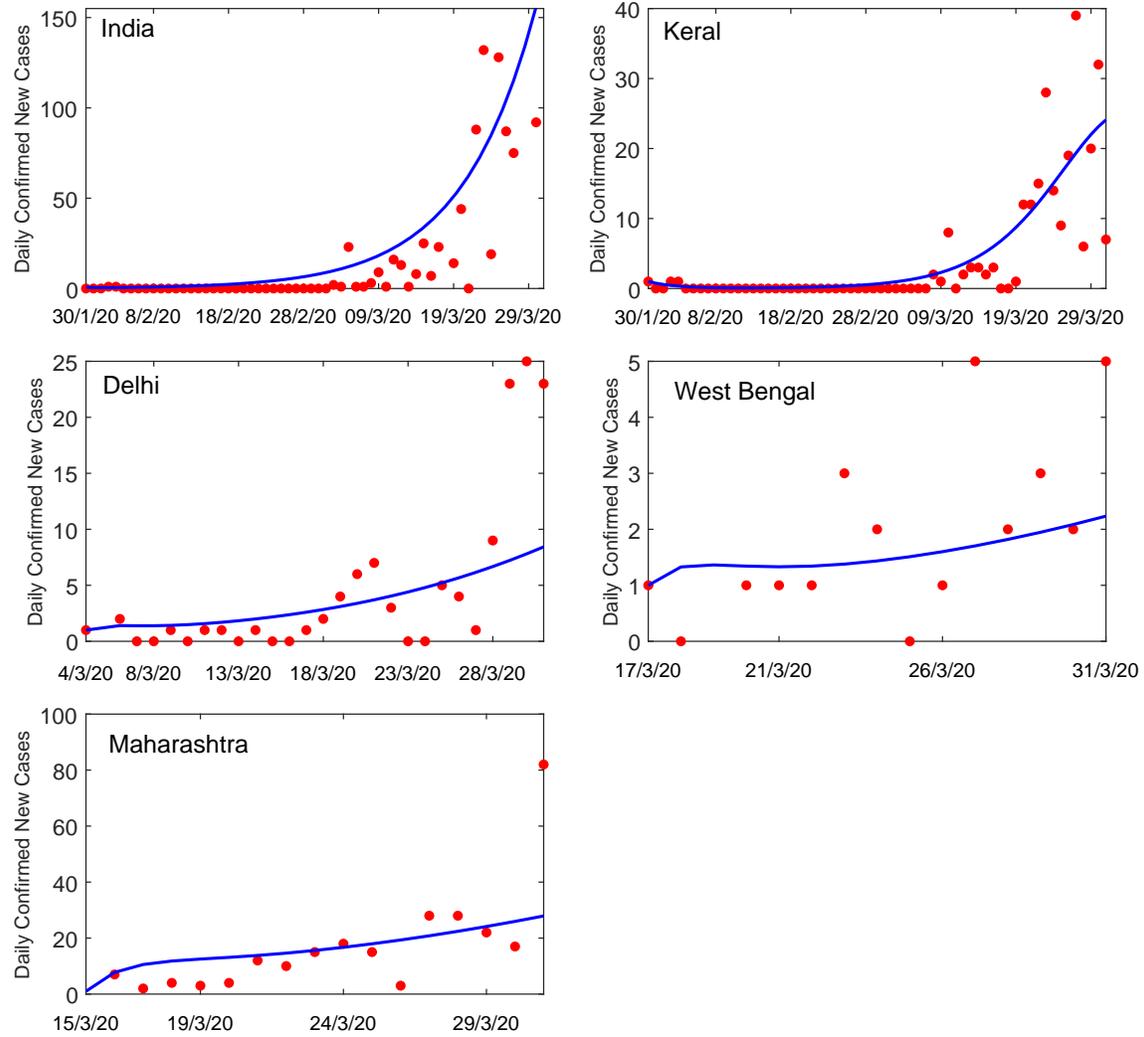}
\caption{Daily new confirmed positive coronavirus cases. Observed data are shown in red circles, whereas the blue curve is the best fitting curve of the model system (\ref{stateeq}), without media awareness.}
\label{DailyNewCases}
\end{figure}

\begin{figure}
\centering
\includegraphics[width=15cm]{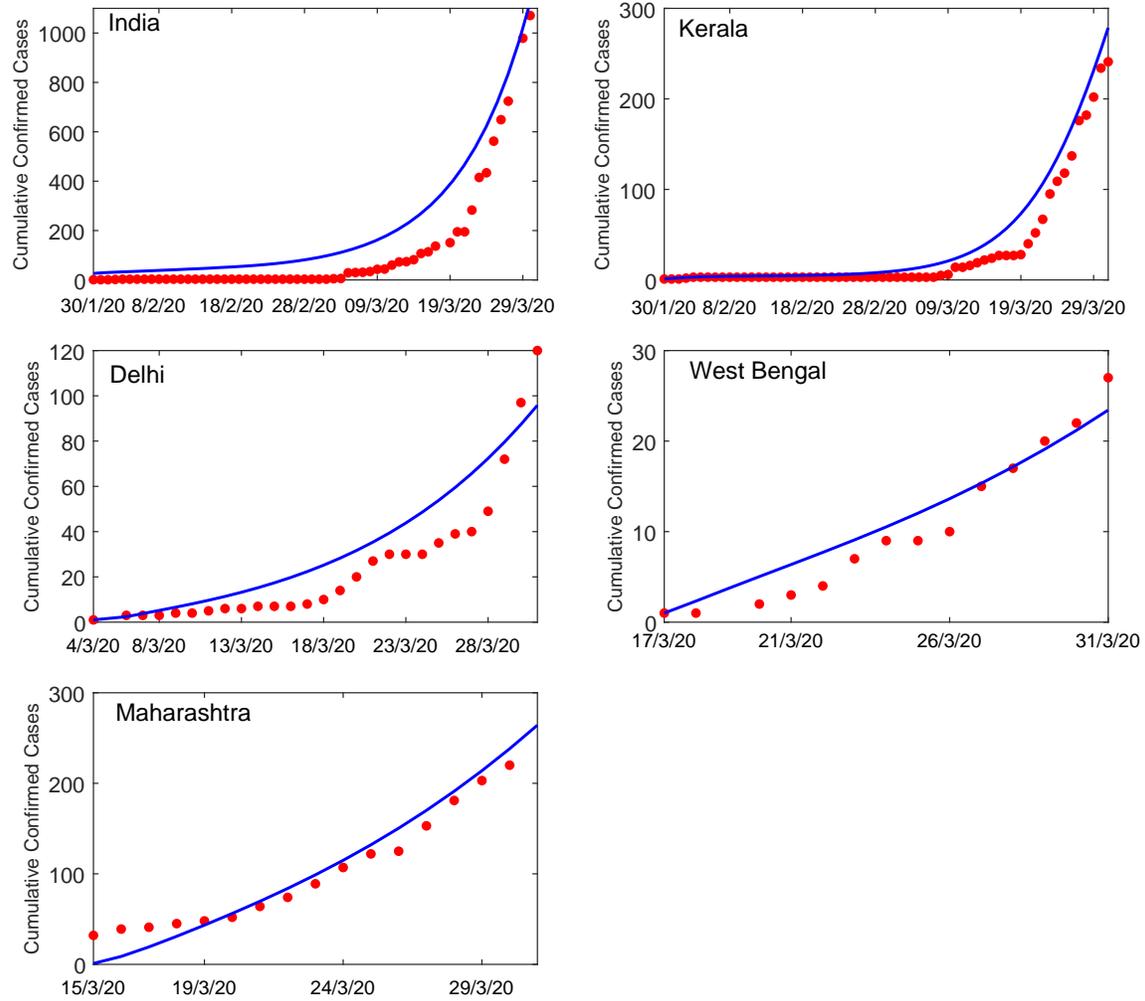}
\caption{Daily cumulative confirmed coronavirus cases. Observed data are shown in red circles, whereas the blue curve is the best fitting curve of the model system (\ref{stateeq}), without media awareness.}
\label{CumulativeCases}
\end{figure}

By fitting our proposed model (simply the classical model without considering media impact) system (\ref{stateeq}) with real data obtained from India COVID-19 Tracker (https://www.covid19india.org/), we were able to get acceptable estimations for the system parameters and curve fitting, as shown in the Figures \ref{DailyNewCases} and \ref{CumulativeCases}. To ensure the performance of our system without media, we compute the numerical errors, namely $E_{MAE}$ and $E_{RMSE}$. Within the reasonable set of parameters, the model simulations without incorporating media awareness, obtained relatively good fits to novel coronavirus data, as plotted in the Figures \ref{DailyNewCases} and \ref{CumulativeCases}.

From the basic reproduction number $R_0$  in Table \ref{parvalR0}, it can be compared that the province Kerala is highly effective by coronavirus diseases compare to other three provinces and the Republic of India. Nevertheless, our model-based analysis also demonstrates that the basic reproduction number in India and four different provinces remains greater than unity. The estimation of basic reproduction number $R_0$ indicates that the strengthen control involvements are mandatory to reduce and/or eradicate the future coronavirus diseases. Thus, in this scenario the public must have to give utmost priority to stay home quarantine and successful the lock-down to combat against coronavirus pandemic. In this situation, media related awareness is mandatory to aware each and every people such that they can maintain social distancing, wear mask to protect themselves and avoid cultural programme.

\begin{figure}
  \centering
  \includegraphics[width=0.9\textwidth]{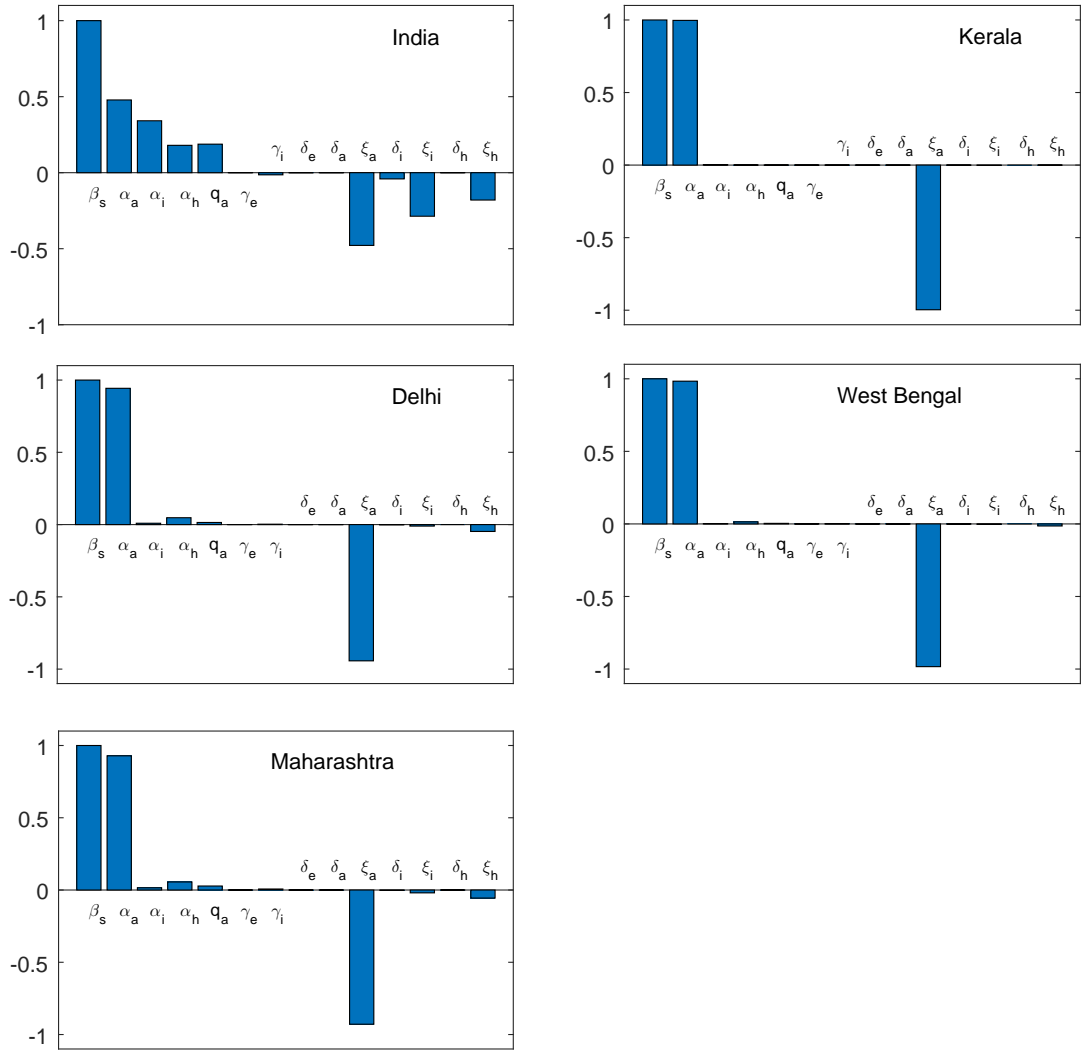} \\
  \caption{ Result shows the normalized forward sensitivity indices of basic reproduction number $R_0$ with respect to each of the baseline parameter values using in the Table \ref{R0_sens}.}
\label{Fsens_r0}
\end{figure}

\begin{figure}
  \centering
  \includegraphics[width=0.9\textwidth]{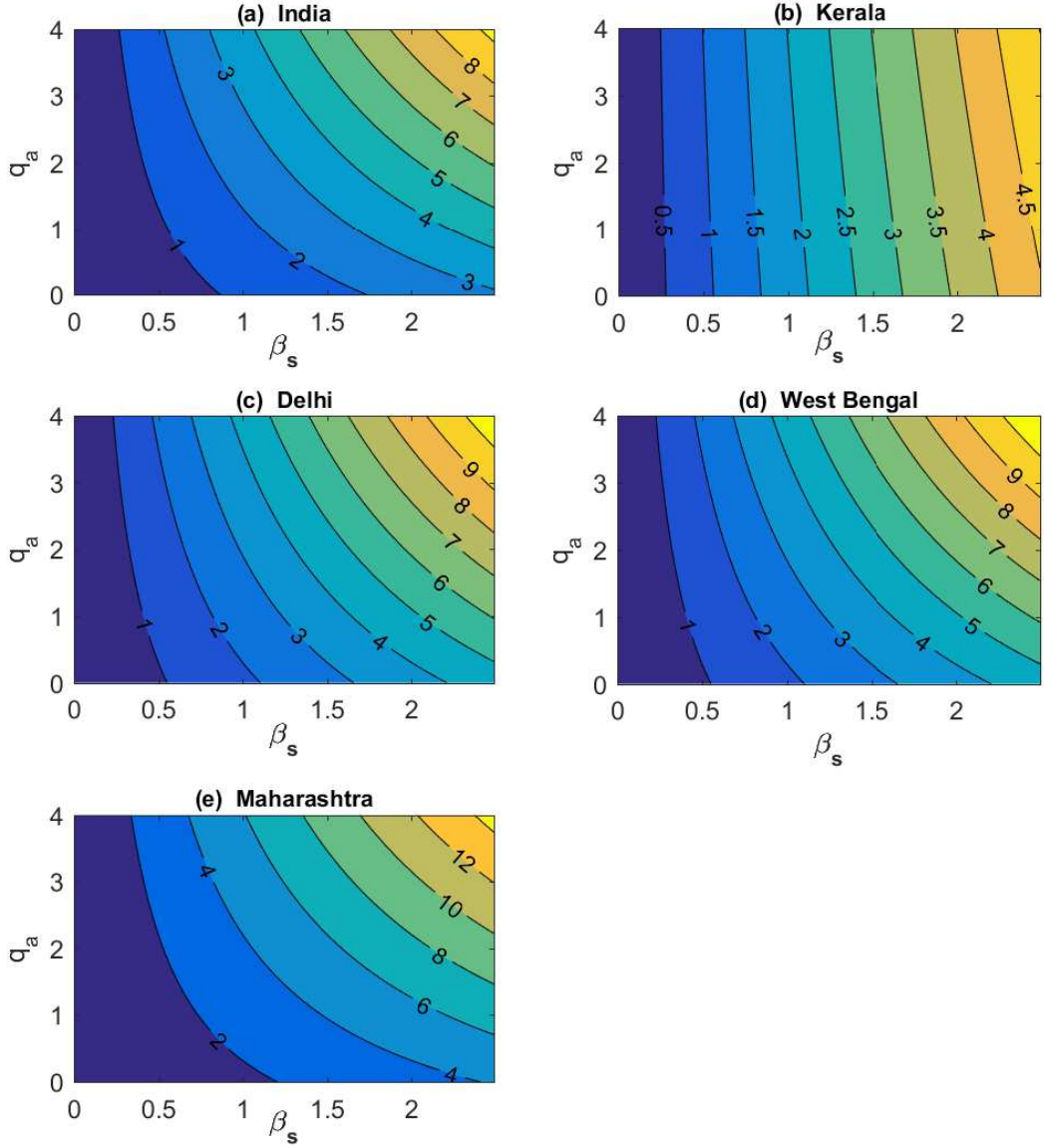} \\
  \caption{ Contour plots of basic reproduction number $R_0$ for the four different provinces and the Republic of India. Plot contours of $R_0$ versus the disease transmission coefficient $\beta_s$ and the portion $q_a$ of exposed class after being clinically ill due to novel coronavirus. For (a) the Republic of India, (b) the province Kerala, (c) Delhi, (d) West Bengal and (e) Maharashtra. The contour plots exhibits that the higher disease transmission probability of coronavirus disease will remarkably increase the basic reproduction number. The baseline parameter values are obtained from the Table \ref{parval} and rest of the parameter values are estimated and obtained from the Table \ref{parvalEst}.}
\label{contour}
\end{figure}

From the sensitivity indices (see Table \ref{R0_sens} and the Figure \ref{Fsens_r0}) for the basic reproduction number $R_0$, we observed that the disease transmission coefficient $\beta_s$ and the portion $q_a$ clinical outbreak rates in all infected classes are positively correlated with $R_0$. This indicates that to reduce coronavirus diseases the social distancing is the most important  factor and we reduce the values for $\beta_s$ and $q_a$ to control the diseases. As we know, for $R_0 < 1$ the diseases can be died out. Thus, to keep control the value of $R_0$ we must reduce $\beta_s$ and $q_a$. While investigating the contour plot, we observe that for the lower values for $\beta_s$ and $q_a$, the reproduction number can be controlled and reduced to less than unity, as shown in the Figure \ref{contour}. Thus, if we are able to control the social distancing, then we are able to control the basic reproduction number $R_0$ and hence we are able to eradicate the coronavirus diseases. Since $R_0$ quantifies the initial epidemic transmission, its sensitivity indices enable us to establish the relative significance of various parameters in coronavirus transmission. From the sensitivity of basic reproduction number $R_0$, we can conclude that the most effective control measures are the isolation of all close contacts and strengthening the self-protection ability for susceptible individuals.

\begin{figure}
\centering
\includegraphics[width=15cm]{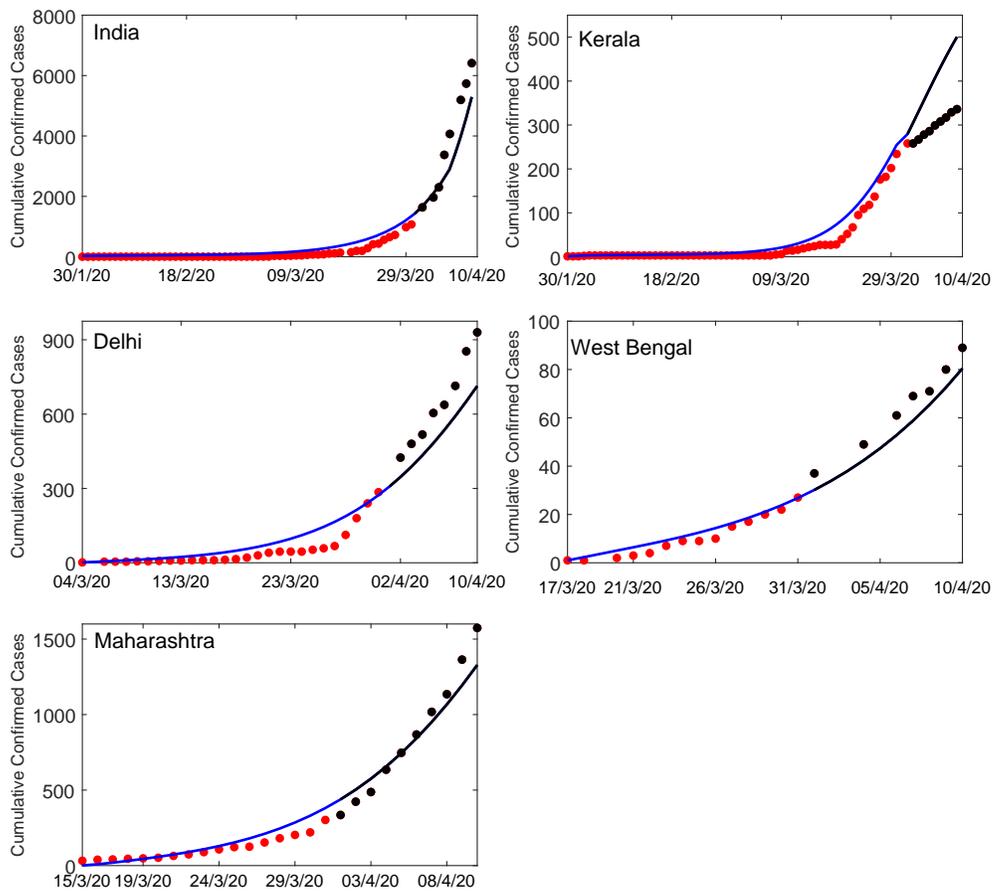}
\caption{Model simulations shows the short term predictions for the four provinces, namely Kerala, Delhi, West Bengal, Maharashtra and the Republic of India. The solid black curve represents the model predictions without incorporating media impact for the new infected cumulative coronavirus cases, whereas the solid black circles are the real data. The baseline parameter values are listed in the Table \ref{parval} and rest of estimated parameters are listed in the Table \ref{parvalEst}. Our model able to capture the increasing tends for three states, namely Delhi, West Bengal, Maharashtra and the Republic of India, of newly infected cumulative coronavirus diseases. In case of Kerala, model performance is slightly difference in compare to other states and the overall India. }
\label{f_pred}
\end{figure}

We have checked our calibrated epidemic model for the short term prediction in the four provinces and the republic of India. The simulation of our calibrated model successfully able to capture the increasing growth patterns for three different provinces, namely Delhi, Maharashtra, West Bengal and the Republic of India, whereas in case of the province Kerala, the model fitting is not good compare to other states and overall India, as plotted in the Figure \ref{f_pred}. To get an answer regarding the question why the model prediction is perfectly matched with Delhi, Maharashtra, West Bengal and the overall India but not effective for Kerala, we have closely look into the estimated parameters (Table \ref{parvalEst}) and come to the conclusion that in Delhi, West Bengal, Maharashtra and overall India there is a large number of symptomatic individuals whereas in Kerala the symptomatic population are decreasing due to effective precautions for coronavirus diseases. Thus, our calibrated model predicted that the increasing pattern of coronavirus diseases will be continued if the people not aware about the diseases. Thus, the public must have to take the precautions and the health care agencies should concentrate on successful implementation of control mechanisms to reduce and/or eradicate the disease.

\begin{figure}
\centering
\includegraphics[width=13cm]{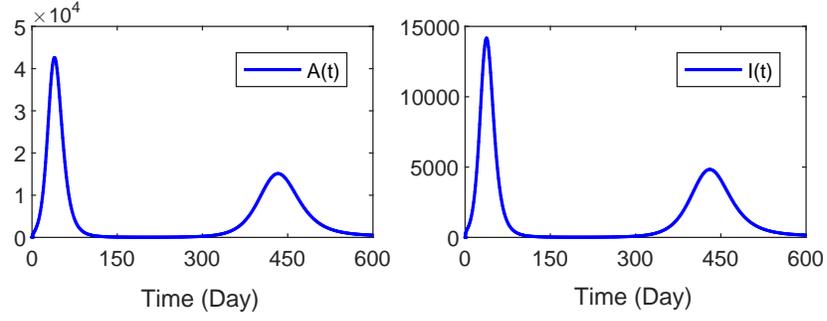}
\caption{Long term dynamics for the coronavirus disease system (\ref{stateeq}) for the infectious individuals (asymptomatic and symptomatic populations). The value of the parameters used for the numerical simulation is given in Table \ref{parval}. The other parameters values are $\beta_s = 1.8326$, $\gamma_i = 0.1423$, $q_a = 0.4110$,$ \xi_a = 0.1738$, $\xi_h = 0.2100$ and $\Lambda_s = 1600$ with the initial individuals $S(0) = 50000$, $E(0) = 5000$, $A(0) = 10$, $I(0) = 1$, $H(0) = 1$, $R(0) = 10$.}
\label{f-longTime}
\end{figure}

In order to study the long-term dynamics of our model, we observed the most interesting switching phenomena, as plotted in the Figure \ref{f-longTime}. Maybe this is happen due to the impact of media awareness as it is a dynamic process on disease epidemic. Also, the media awareness has a good impact in mitigating the disease transmission at an initial stage of an epidemic. It is worthy to mention that the media impact is not always remain efficacious for mitigating disease transmission throughout the epidemic \cite{Collinson14}. A detailed understanding of media awareness during an outbreak can help in the progression of an implementable public health strategy. A particular interest to the designers of such strategies are the effects of media awareness on some significant epidemic features such as the size of the peak, its timing and the whole number of diseases. Our proposed model and its simulation, like those in earlier investigations \cite{Buonomo08,Collinson14,d'Onofrio07,d'Onofrio09,Funk09,Funk10,Kumar17,Reluga06,Sun11,Das20,Xiao15}, allow that quantifying these media effects gives additional insights.

Our model simulation and prediction suggests that the novel coronavirus diseases has a potentiality to exhibit oscillatory dynamics in the future but can be controllable by maintaining social distances and effectiveness of isolation or hospitalization. Our model forecasts that isolation or hospitalization of populations with coronavirus symptoms, under stringent hygiene safeguards and social distancing, be able to effectual control in a community and may even eliminate the diseases. Our study also suggests that the size and duration of an epidemic can be considerably affected by timely implementation of the hospitalization or isolation programme. Based on the model simulations and prediction to control the coronavirus diseases, we provide some particular inferences to face with emerging diseases:
\begin{description}
  \item[i.] avoidance of mass gatherings, rallies, social distancing and implement extensive lock-down;
  \item[ii.] provide essential personal safeguard to the staff who are involved in emergency services;
  \item[iii.] timely give the statistics of the coronavirus diseases to the community, including the number of asymptomatic cases, symptomatic cases, hospitalized cases and so on;
  \item[iv.] increase the media awareness to aware the population, organize the targeted health care education and self-protection and remove the public mental disorder.
\end{description}

\clearpage

\noindent \\ \textbf{Data availability} \\
All data supporting the findings of this study are in the paper and available from the corresponding author on reasonable request.

\noindent \\ \textbf{Author contributions statement} \\
Subhas Khajanchi, Kankan Sarkar, and Jayanta Mondal designed and performed the research as well as wrote the paper.

\noindent \\ \textbf{Competing interests} \\
The authors declare that they have no conflict of interest.

\clearpage

\clearpage

\clearpage

\clearpage

\clearpage

\clearpage

\clearpage

\clearpage

\clearpage

\clearpage

\clearpage

\begin{table}[ht]
  \caption{Table of biologically relevant parameter values and their description for the coronavirus model system (\ref{stateeq}).}
\label{parval}
 {\begin{tabular}{l l l l l }
  \\[-0.1cm]
  \hline
  Parameter  & Biological meaning & Values (Unit) & Source   \\[0.01cm]
  \hline
 $\Lambda_s$ & Inflow rate of susceptible individual & - & - \\
  $\beta_s$  & Disease transmission coefficient & -  & Estimated \\
  $\alpha_a$ & Adjustment factor for asymptomatic classes & 0.20  ~$day^{-1}$ & \cite{Gumel04} \\
  $\alpha_i$ & Adjustment factor for symptomatic classes & 0.45 ~$day^{-1}$ & \cite{Nadim20} \\
  $\alpha_h$ & Adjustment factor for hospitalized individuals & - & Estimated  \\
  $\delta_s$ & Natural death rate of susceptible classes & 0.1945 $\times 10^{-4}$ ~$day^{-1}$ & \cite{data-gov} \\
  $\delta_e$ & Mortality rate of exposed individuals & 0.1945 $\times 10^{-4}$ ~$day^{-1}$ & \cite{data-gov} \\
  $\delta_a$ & Mortality rate of asymptomatic individuals & 0.1945 $\times 10^{-4}$ ~$day^{-1}$ & \cite{data-gov} \\
  $\delta_i$ & Mortality rate of symptomatic individuals & 0.03 ~$day^{-1}$ & \cite{Who-46} \\
  $\delta_h$ & Mortality rate of hospitalized individuals & 0.1945 $\times 10^{-4}$ ~$day^{-1}$ & \cite{data-gov} \\
  $\delta_r$ & Mortality rate of recovered individuals & 0.1945 $\times 10^{-4}$ ~$day^{-1}$ & \cite{data-gov} \\
  $\gamma_e$ & Conversion rate from exposed to asymptomatic  individuals & 1/7 ~$day^{-1}$ & \cite{A-who} \\
  $\gamma_i$ & Rate at which symptomatic individuals become hospitalized & -  & Estimated \\
  $q_a$      & Proportion of exposed individuals & - & Estimated \\
  $\xi_a$    & Rate of recovery from asymptomatic individuals & -  & Estimated  \\
  $\xi_i$    & Rate of recovery from symptomatic individuals & - & Estimated \\
  $\xi_h$    & Rate of recovery from hospitalized individuals & -  & Estimated  \\
  $\delta_m$ & Degradation of media related information & 0.06  & \cite{Kumar17} \\
  $\eta_m$   & Rate of media response & (0, ~1)  & \cite{Xiao15} \\
  $d_m$      & Media related information index rate & 0.017 ~$day^{-1}$  & \cite{Kumar17} \\
  $\alpha_m$ & Source rate of the information index & (0, ~1)  & \cite{Buonomo08,d'Onofrio07} \\
  \hline
  \end{tabular}}
\end{table}

\clearpage

\begin{table}[h]
\caption{\emph{Estimated parameter values from the observed data.}}
\begin{center}
 \begin{tabular}{p{2.0cm} p{1.8cm} p{1.8cm} p{1.8cm} p{2.2cm} l}
 \hline
Parameters & India  & Kerala & Delhi   & West Bengal & Maharashtra \\
 \hline
$\beta_s$ & 1.8326  & 2.6822 & 1.7880 & 1.5013 & 1.7781\\
$\xi_a $  & 0.1738 & 0.1121 & 0.1105 & 0.1101 & 0.1203\\
$q_a$     & 0.4110 & 0.0028 & 0.0430 & 0.0125 & 0.0448\\
$\xi_h$   & 0.2100 & 0.2137 & 0.1358 & 0.1332 & 0.1124\\
$\gamma_i$& 0.1423 & 0.0348 & 0.8900 & 0.8953 & 0.5213\\
$\xi_i$ & 0.2100 & 0.2105 & 0.2091 & 0.2097 & 0.2109\\
$\alpha_h$ & 0.3500 & 0.3491 & 0.3506 & 0.3489 & 0.3501\\
 \hline
\end{tabular}
\end{center}
\label{parvalEst}
\end{table}

\clearpage

\begin{table}[h]
\caption{\emph{Accuracy of the model (\ref{stateeq}) for the four different states and the Republic of India.}}
\begin{center}
\begin{tabular}{p{3.5cm} p{1.6cm} p{1.6cm} p{1.6cm} p{2.2cm} l}
 \hline
Performance Metrics & India  & Kerala  & Delhi   & West Bengal & Maharashtra\\
 \hline
$E_{MAE}$ & 10.0600 & 2.4286 & 3.3326 & 1.0294 & 8.7160\\
$E_{RMSE}$ & 17.6314 & 4.9185 & 5.8537 & 1.4416 & 15.4680\\
  \hline
\end{tabular}
\end{center}
\label{error}
\end{table}

\clearpage

\begin{table}[h]
\caption{\emph{Estimated values values of the initial population size and the constant inflow rate}}
\begin{center}
\begin{tabular}{p{2.0cm} p{1.6cm} p{1.6cm} p{1.6cm} p{2.2cm} p{2.0cm} l}
 \hline
Parameters & India  & Kerala & Delhi  & West Bengal & Maharashtra & Source\\
 \hline
$S(0)$ & 50000 & 40000 & 2500 & 12000 & 15000 & Estimated\\
$E(0)$ & 0 & 8 & 300 & 1000 & 1700 & Estimated\\
$A(0)$ & 0 & 1 & 13 & 100 & 250 & Estimated\\
$I(0)$ & 1 & 1 & 1 & 1 & 1 & Data\\
$H(0)$ & 1 & 1 & 1 & 1 & 1 & Estimated\\
$R(0)$ & 0 & 0 & 0 & 0 & 0 & Data\\
$M(0)$ & 0 & 0 & 0 & 0 & 0 & Estimated\\
$\Lambda_s$ & 1600 & 500 & 500 & 1100 & 1100 & Estimated\\
  \hline
\end{tabular}
\end{center}
\label{parvalIC}
\end{table}

\clearpage

\begin{table}[h]
\caption{\emph{Basic reproduction number of the four different provinces and the Republic of India.}}
\begin{center}
\begin{tabular}{p{4.5cm} p{1.6cm} p{1.6cm} p{1.6cm} p{2.2cm} l}
 \hline
Basic reproduction number & India  & Kerala  & Delhi & West Bengal & Maharashtra\\
 \hline
$R_0$ & 2.58792 & 4.78435 & 3.28271 & 2.73854 & 3.03971\\
  \hline
\end{tabular}
\end{center}
\label{parvalR0}
\end{table}

\clearpage

\begin{table}[h]
\caption{\emph{Table of sensitivity indices of basic reproduction number $R_0$ for the coronavirus model system (\ref{stateeq}) for four different states and the Republic of India. The baseline parameter values taken from the Table \ref{parval} and other parameters are estimated and taken from the Table \ref{parvalEst}.}}
\begin{center}
\begin{tabular}{l l l l l l}
 \hline
Parameters & & &Sensitivity & Indices & \\
 \hline
 & India  & Kerala  & Delhi   & West Bengal & Maharashtra\\
 \hline
$\beta_s$  & 1.0000   & 1.0000 & 1.0000& 1.0000& 1.0000\\
$\alpha_a$ & 0.478462 & 0.99711 & 0.943025& 0.983684& 0.928715\\
$\alpha_i$ & 0.341547 & 0.00256553& 0.00933187& 0.0021075& 0.0154709\\
$\alpha_h$ & 0.179991 & 0.000324077& 0.0476427& 0.0142081& 0.0558138\\
$q_a$      & 0.187671 & 0.0000898542& 0.0146025& 0.0038639& 0.0277269\\
$\gamma_e$ & 0.000136131& 0.000136131& 0.000136131& 0.000136131& 0.000136131\\
$\gamma_i$ & -0.0141361& -0.0000411902& 0.00273315& 0.00133818& 0.00705925\\
$\delta_e$ & -0.000136131& -0.000136131& -0.000136131& -0.000136131& -0.000136131\\
$\delta_a$ & -0.0000535388& -0.000172974& -0.00016596& -0.000173745& -0.00015013\\
$\xi_a $   & -0.478408& -0.996937& -0.942859& -0.983511& -0.928565\\
$\delta_i$ & -0.0409264& -0.000314886& -0.0015138& -0.000431249& -0.00280575\\
$\xi_i$    & -0.286484& -0.00220945& -0.0105512& -0.00301443& -0.0197244\\
$\delta_h$ & -0.0000166691& -2.94934$\times 10^{-8}$& -6.82266$\times 10^{-6}$&-2.07434$\times 10^{-6}$& -9.6565$\times 10^{-6}$\\
$\xi_h$    & -0.179975& -0.000324048& -0.0476359& -0.014206& -0.0558041\\
    \hline
\end{tabular}
\end{center}
\label{R0_sens}
\end{table}

\end{document}